\documentclass[12pt,preprint]{aastex}
\usepackage{graphicx}
\usepackage[T1]{fontenc}
\shorttitle{V-type asteroids in the Outer Belt}
\shortauthors{Duffard and Roig}
\begin{document}
\title{Two new V-type asteroids in the outer Main Belt?\altaffilmark{1}}
\altaffiltext{1}{Based on observations collected at the Centro
Astron\'{o}mico Hispano Alem\'{a}n (CAHA) at Calar Alto, operated
jointly by the Max-Planck Institut f\"{u}r Astronomie and the Instituto
de Astrof\'{\i}sica de Andaluc\'{\i}a (CSIC).}
\author{R. Duffard\altaffilmark{*2}}
\affil{Instituto de Astrof\'{\i}sica de Andaluc\'{\i}a,
C/ Bajo de Huetor, 50. 18008. Granada, Spain, and\\
Max Planck Institute for Solar System Research, Max Planck Str. 2,
Katlenburg-Lindau, Germany} \email{duffard@iaa.es}
\altaffiltext{2}{corresponding author} \and
\author{F. Roig}
\affil{Observat\'{o}rio Nacional, Brazil}
\email{froig@on.br}
\begin{abstract}
The identification of basaltic asteroids in the asteroid Main Belt
and the description of their surface mineralogy is necessary to
understand the diversity in the collection of basaltic meteorites.
Basaltic asteroids can be identified from their visible
reflectance spectra and are classified as V-type in the usual
taxonomies. In this work, we report visible spectroscopic
observations of two candidate V-type asteroids, (7472) Kumakiri
and (10537) 1991 RY16, located in the outer Main Belt ($a>2.85$ UA). These
candidate have been previously identified by Roig and Gil-Hutton
(2006, Icarus 183, 411) using the Sloan Digital Sky Survey colors.
The spectroscopic observations have been obtained at the Calar
Alto Observatory, Spain, during observational runs in November
and December 2006. The spectra of these two asteroids show the steep slope
shortwards of 0.70 $\mu$m and the deep absorption feature
longwards of 0.75 $\mu$m that are characteristic of V-type
asteroids. However, the presence of a shallow but conspicuous absorption band
around 0.65 $\mu$m opens some questions about the actual mineralogy
of these two asteroids. Such band has never been observed before
in basaltic asteroids with the intensity we detected it. We discuss the possibility
for this shallow absorption feature to be caused by the presence
of chromium on the asteroid surface. Our results indicate that,
together with (1459) Magnya, asteroids (7472) Kumakiri and (10537)
1991 RY16 may be the only traces of basaltic material found
up to now in the outer Main Belt.
\end{abstract}
\keywords{Asteroids, composition}
\section{Introduction}
Basaltic asteroids are small bodies connected to the processes of heating and melting
that may have led to the mineralogical differentiation in the interiors of the largest
asteroids. Therefore, a precise knowledge of the inventory of basaltic asteroids
may help to estimate how many differentiated bodies actually formed
in the asteroid Main Belt, and this in turn may provide important constraints to the
primordial conditions of the solar nebula.

In the visible wavelengths range, the reflectance spectrum of
basaltic asteroids is characterized by a steep slope shortwards of
0.70 $\mu$m and a deep absorption band longwards of 0.75 $\mu$m.
Asteroids showing this spectrum are classified as V-type in the
usual taxonomies (e.g. Bus \& Binzel, 2002).

A few years ago, most of the known V-type asteroids were members of the
Vesta dynamical family, located in the inner asteroid belt
--semi-major axis $a < 2.5$ AU--. This family
formed by the excavation of a large crater (Thomas et al., 1997; Asphaug, 1997)
on the surface of asteroid (4) Vesta, which is the only known large
asteroid --diameter $D\sim 500$ km-- to show a basaltic crust (McCord et al., 1970).

Nowadays, however, several V-type asteroids have been identified in the inner belt
but outside the Vesta dynamical family (Burbine et al., 2001; Florczak et al., 2002;
Alvarez-Candal et al., 2006). Basaltic asteroids have also been found
in the middle Main Belt --$2.5 < a < 2.8$ AU-- (Binzel et al., 2006; Roig et al., 2007), as well
as among the Near Earth Asteroids (NEA) population (McFadden et al., 1985; Cruikshank et al., 1991;
Binzel et al., 2004; Duffard et al., 2006). Recent works (Carruba et al., 2005, 2007;
Nesvorn\'{y} et al., 2007; Roig et al., 2007) provide evidence that many of these
V-type asteroids may be former members of the Vesta family, that reached their
present orbits due to long term dynamical evolution. The exception is asteroid
(1459) Magnya, the only basaltic object so far discovered in the outer belt
--$a > 2.8$ AU-- (Lazzaro et al., 2000). This asteroid is too far away from the Vesta family
and it is also too big --$D = 20$-30 km-- to
have a real probability of being a fragment from the Vesta's crust
(Michtchenko et al., 2002).

Beyond the existence of (4) Vesta and the V-type asteroids related
to the Vesta dynamical family, the paucity of intact
differentiated asteroids and of their fragments observed today in
the main belt is an strong constraint to the formation scenario of
basaltic material. The sample of iron meteorites collected in the
Earth indicates that they would come from the iron core of dozens
of differentiated parent bodies. However, there are very few
olivine-rich asteroids (classified as A-type) which are assumed to
come from the mantle of differentiated bodies, and only one
asteroid, (1459) Magnya, is known to sample the basaltic crust of
a differentiated parent body other than (4) Vesta. Finally, the
other Main Belt asteroid families, which formed from the
disruption of over fifty $10<D<400$ km asteroids, show little
spectroscopic evidence that their parent bodies were heated enough
to produce a distinct core, mantle and crust (Cellino, 2003).

Aiming to establish if other V-type asteroids might be found together with (1459) Magnya in
the outer belt, thus giving support to the existence of a differentiated parent body in
that part of the belt, Roig \& Gil-Hutton (2006) used the five band photometry from the
3rd release of the Sloan Digital Sky Survey Moving Objects Catalog
(SDSS-MOC; Ivezi\'c et al., 2001; Juri\'c et al., 2002) to identify
candidate V-type asteroids. Among 263 candidates that are not members
of the Vesta dynamical family, Roig \& Gil-Hutton found
five possible V-type asteroids in the outer belt: (7472) Kumakiri, (10537) 1991 RY16,
(44496) 1998 XM5, (55613) 2002 TY49, and (105041) 2000 KO41.
However, these findings need to be confirmed by accurate spectroscopic
observations.

The aim of this work is to describe the visible spectroscopic
observations of two of these candidates: (7472) Kumakiri and
(10537) 1991 RY16. Our goal is to provide a more reliable
taxonomic classification of these asteroids indicating that
\textit{they would the second and third basaltic asteroids
discovered up to now in the outer belt}. Our observations also
reveal certain peculiarities of their spectra that deserve special
attention in future studies. Last but not least, our results help
to validate the approach of Roig \& Gil-Hutton (2006) to predict
V-type asteroids. It is worth recalling that a similar study has
been performed by Roig et al. (2007), who used visible
spectroscopic observations taken at the Gemini Observatory to
confirm the classification of two candidate V-type asteroids in
the middle belt: (21238) 1995 WV7 and (40521) 1999 RL95.

In Sect. 2, we describe the observations and the reduction procedures. In Sect. 3,
we present and discuss the results obtained. Finally, Sect. 4 is devoted to conclusions.

\section{Observations}

Low resolution spectroscopy of (7472) Kumakiri and (10537) 1991
RY16 were obtained on November 14, 2006, as part of a 4 nights
observational run, using the Calar Alto Faint Object Spectrograph
(CAFOS) at the 2.2m telescope in Calar Alto Observatory, Spain.
The prime aim of the run was to characterize V-type asteroids
inside and outside the Vesta family. Asteroid (7472) Kumakiri was
observed again on December 29, 2006, using the same
instrument and telescope, under Director's Discretionary Time (DDT).
Table \ref{observ} summarizes the observational circumstances.

CAFOS\footnote{%
See \texttt{http://www.caha.es/alises/cafos/cafos22.pdf} for more
details.} is equipped with a 2048$\times$2048 CCD detector SITe-1d
(pixel size 24 $\mu$m/pixel, plate scale 0.53"/pixel). We used the
R400 grism allowing to obtain an observable spectral range between
0.50 and 0.92 $\mu$m. To remove the solar component of the spectra
and obtain the reflectance spectra, the solar analog stars HD
191854, HD 20630 and HD 28099 (Hardorp, 1978) were also observed
at similar airmasses as the asteroids. In order to estimate the
quality of each night, at least two solar analogs were observed
per night and we verified that the ratios between the
corresponding spectra show no significant variations. Bias frames,
spectral dome flat fields and calibration lamps spectra were also
taken in each night to allow reduction of the science images.
Spectrum exposures for each asteroid were splitted in two
exposures at two different slit positions, A and B, separated by
20" (the width of the slit was 2.0"). The observations were
performed with the telescope tracking at the proper motion of the
asteroid. Hence by subtracting A from B and B from A, a very
accurate background removal is achieved. Finally, standard methods
for spectra extraction were applied.

%________________________________________________________________
%
\section{Results and discussion}

\begin{figure}
\centering
\includegraphics[width=13cm]{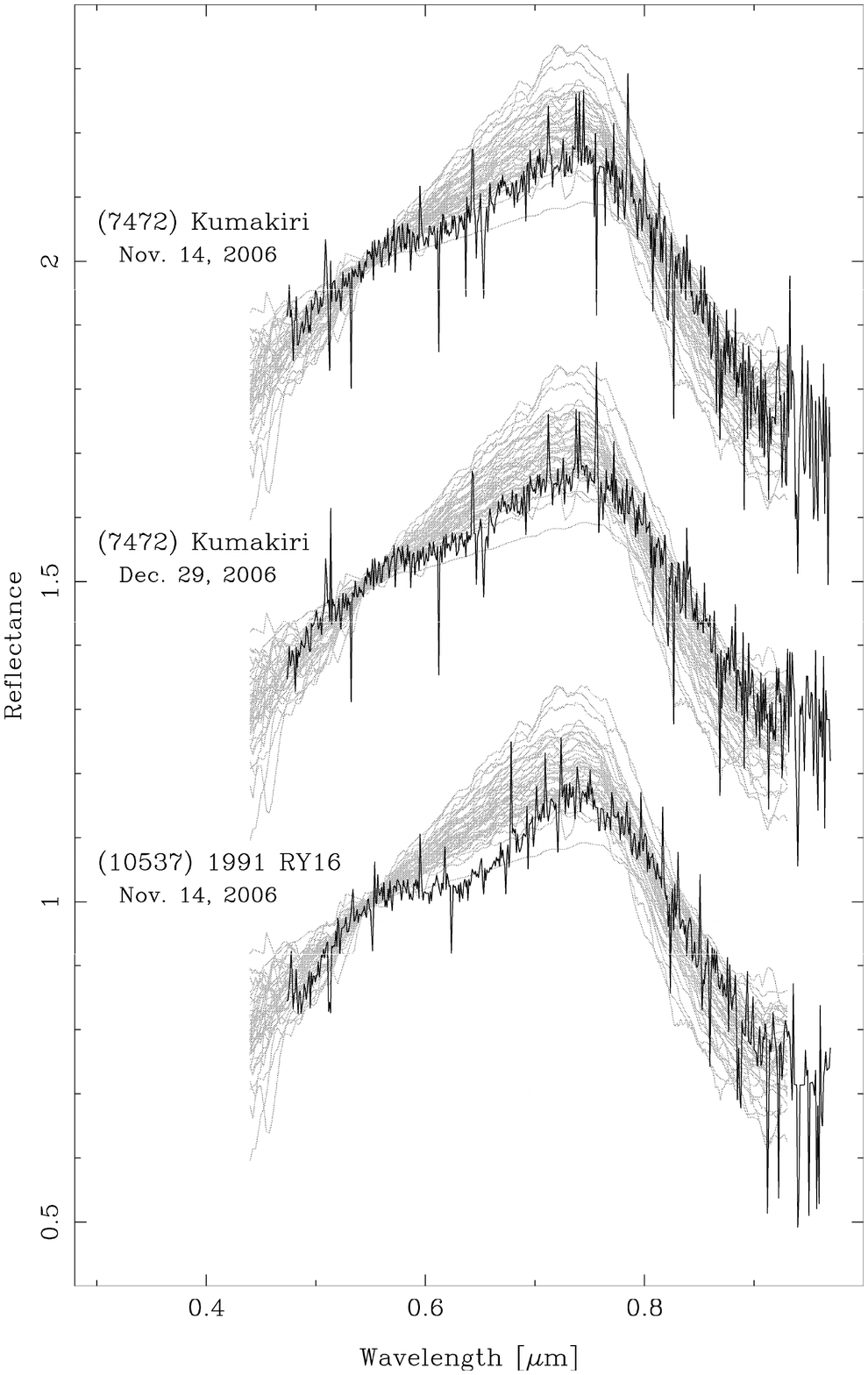}
\caption{Reflectance spectra of (7472) Kumakiri and (10537) 1991
RY16 (black lines) compared to the spectra of several known V-type
asteroids taken from the SMASS and S3OS2 surveys (gray lines). The
spectra are normalized to 1 at 0.55 $\mu$m and shifted by 0.5
units in reflectance for clarity. To remove the solar
contribution, we have used the solar analog HD 191854 in the
November 14 observations and the solar analog HD 28099 in the
December 29 observation.} \label{fig1}
\end{figure}

\begin{figure}
\centering
\includegraphics[width=13cm]{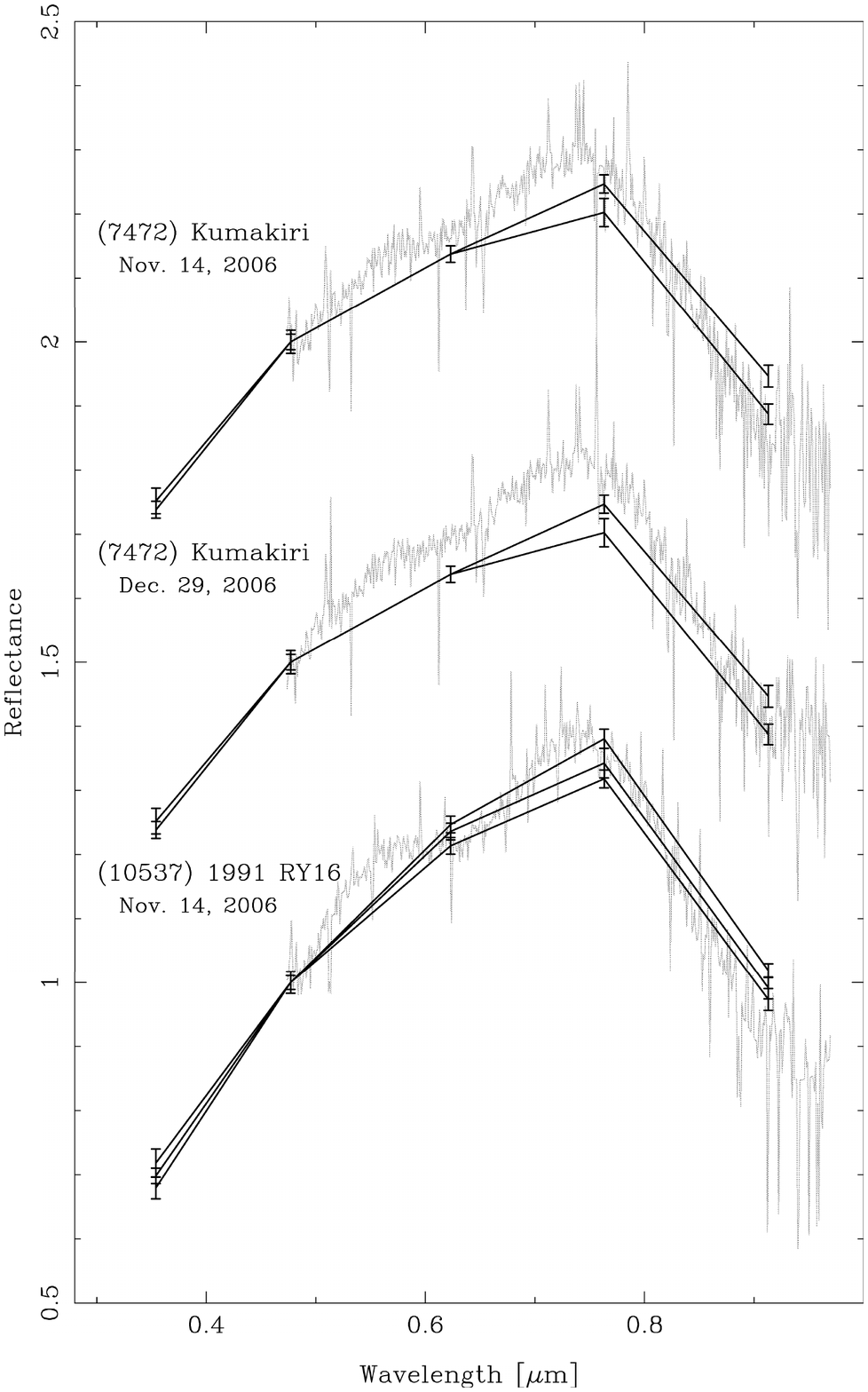}
\caption{Reflectance spectra of (7472) Kumakiri and (10537) 1991
RY16 (gray lines) compared to the photometric observations of the
SDSS-MOC (black lines). The spectra are normalized to 1 at 0.477
$\mu$m (i.e. the center of the $g$ band in the SDSS photometric
system), and shifted by 0.5 units in reflectance for clarity. The
errors in the SDSS-MOC fluxes are less than 3\%.} \label{fig2}
\end{figure}

The reflectance spectra of (7472) Kumakiri and (10537) 1991 RY16
are shown in Figs. \ref{fig1} and \ref{fig2}. Both spectra show a
steep slope shortwards of 0.70 $\mu$m and a deep absorption band
longwards of 0.75 $\mu$m. Using the algorithm of Bus (1999), we
determine that the spectra can be classified as V-type. Figure
\ref{fig1} show that our observations are compatible with the
spectra of previously known V-type asteroids (gray lines) taken
from the SMASS survey (Bus \& Binzel, 2002) and the S3OS2 survey
(Lazzaro et al., 2004). Figure \ref{fig2} show the good agreement
between the five band photometry of the SDSS-MOC (black lines) and
the observed spectra. It is worth noting that the values of
maximum and minimum reflectance prevents to attribute to these
spectra other taxonomic classification, like R-, O- or Q-type. In
view of this, (7472) Kumakiri and (10537) 1991 RY16 may be
considered, together with (1459) Magnya, the only V-type asteroids
discovered up to now in the outer belt.

Notwithstanding, the spectra of (7472) Kumakiri and (10537) 1991
RY16 show a shallow absorption feature around 0.60-0.70 $\mu$m
that has never been reported before in V-type asteroids. This
feature is more evident in the spectrum of (10537) 1991 RY16.
After the its identification in the November 14 observations, and
excluding possible reduction artifacts or solar analogs problems,
we requested Director Discretionary Time (DDT) for another
observational run on December 29. Only the spectra of (7472)
Kumakiri was able to be observed during this run, confirming the
presence of the absorption band. Nevertheless, the band in the
spectrum of (10537) 1991 RY16 has also been observed independently
by Moskovitz et al. (2007).

\begin{figure}
\centering
\includegraphics[width=\columnwidth]{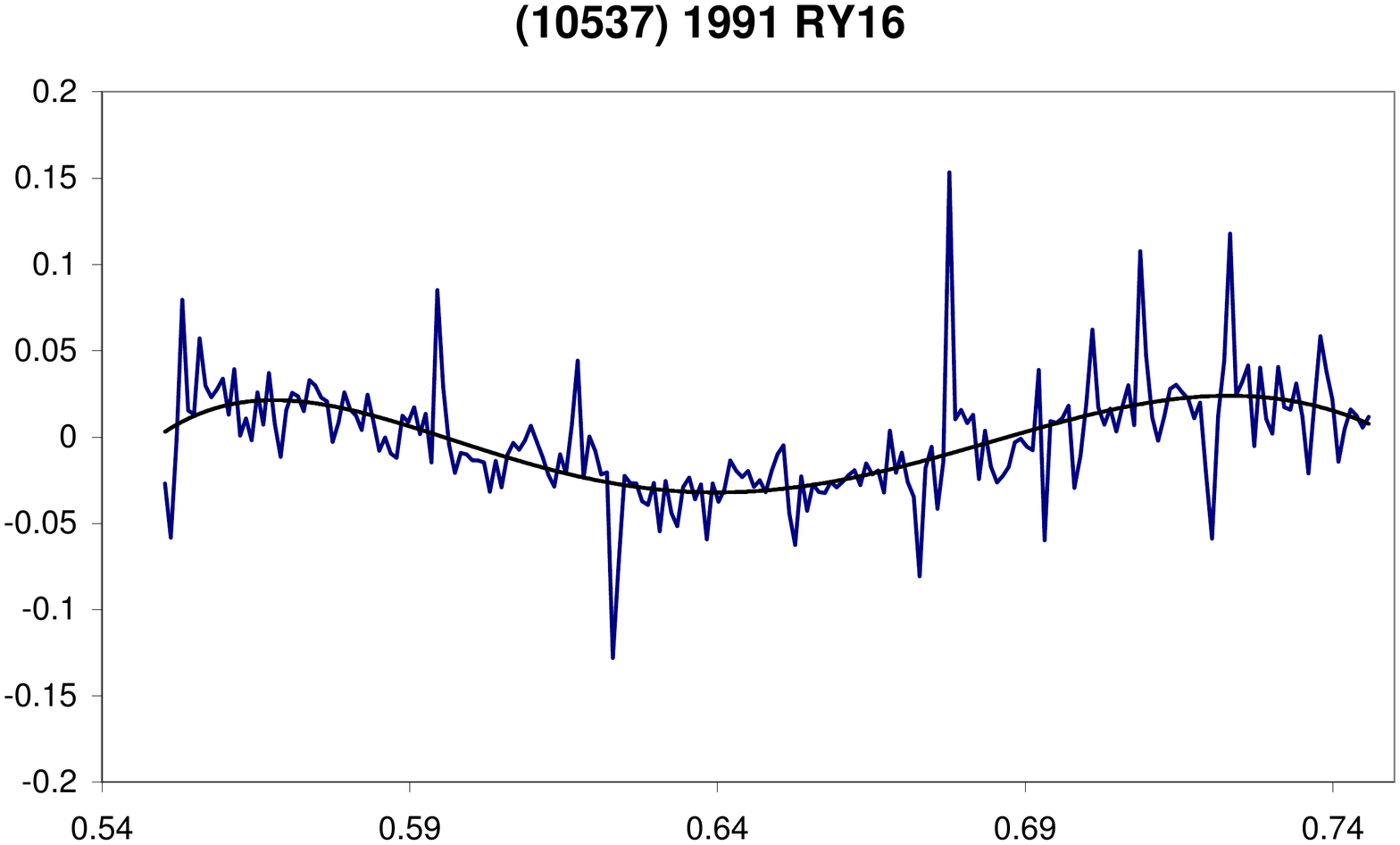}
\caption{Reflectance spectrum of (10537) 1991 RY16 in the
0.55-0.75 $\mu$m interval. The spectrum has been rectified by
subtracting a linear continuum in this interval. From a polynomial
fit (thick line), the center of the absorption band is detected at
0.63 $\mu$m with a FWHM of 0.1 $\mu$m. } \label{fig3}
\end{figure}

To analyze this band, we rectified the spectra by subtracting a
linear continuum in the interval 0.55 and 0.75 $\mu$m and then
fitted several polynomials of different degrees. This allowed to
determine the center of the band at $0.63 \pm 0.01$ $\mu$m and the
FWHM of $\sim 0.1$ $\mu$m (e.g. Fig. \ref{fig3}).

The origin of this absorption band is unclear. Such kind of bands are
usually believed to arise from the $\mathrm{Fe}^{2+}\rightarrow
\mathrm{Fe}^{3+}$ charge transfer absorptions in phyllosilicate
(hydrated) minerals (Vilas \& Gaffey, 1989; Vilas et al., 1993).
However, it is difficult to explain the presence of a hydrated
mineral in the surface of a basaltic object, because the heating
and melting that produce the basalt also eliminate any traces of
water.

It is known that pyroxene crystals Fe$^{2+}$ cations do not show
any absorption bands in the spectral region from 0.56 to 0.72
$\mu$m. Therefore, the origin of the observed band might be
related to other impurity cations like Mn$^{2+}$, Cr$^{3+}$, and
Fe$^{3+}$, usually located in the M1 site of terrestrial and
meteorite orthopyroxenes (Shestopalov et al., 2007). In
particular, broad spin-allowed bands of trivalent chromium around
0.430-0.455 $\mu$m and 0.620-0.650 $\mu$m have been observed in
both reflectance and transmitted spectra of Cr-containing
terrestrial ortho and clinopyroxenes (see Cloutis, 2002), as well
as in diogenite reflectance spectra (McFadden et al., 1982).
Cr$^{3+}$ cations also give spin-forbidden bands near 0.480,
0.635, 0.655, and 0.670 $\mu$m but they do not give absorptions
near 0.57 $\mu$m.

Cloutis (2002) specifically found that Cr$^{3+}$ gives rise to an
absorption band near 0.455 $\mu$m and a more complex absorption
feature in the 0.65 $\mu$m region. However, changes in the grain
size of the pyroxenes may have an effect on the depth of these
absorption bands (Cloutis and Gaffey 1991; Sunshine and Pieters
1993). Therefore, the presence of specific absorption bands can be
taken as an evidence for the presence of a particular cation, but
the characteristics of these bands (depth and width) are probably
not reliable enough to constrain the cation abundance (Cloutis
2002). For example, the grain size may be responsible of the
different band depth observed between the spectra of (7472)
Kumakiri and (10537) 1991 RY16. The slight differences in the band
profile between the November and December spectra of (7472)
Kumakiri might be attributed to different rotational phases\footnote{%
We have verified that these differences cannot be related to
observation/reduction problems, since we do not find any
differences between the spectra of the solar analog stars used in
the different nights.}.

Another interesting feature observed in our spectra is that the
band center of the major absorption feature at 0.90 $\mu$m is
displaced to larger wavelenghts. In our spectra, this region is
the noisiest but using different polynomial fits it was possible
to estimate the center of the band nearer to 0.92-0.93 $\mu$m.
This behavior may also be attributed to the presence of chromium
on the surface. Actually, Cloutis and Gaffey (1991) suggested that
the Cr-rich pyroxene samples in their study have the two major
absorption features (i.e. the one centered at 0.9 and the one
centered at 1.9 $\mu$m, respectively) displaced to larger
wavelengths than expected, relative to their Fe contents. These
authors also presented the predicted {\it versus} actual
wavelength position of the major Fe$^{2+}$ absorption band center
in the 1 $\mu$m region, and this center is closer to 0.92 $\mu$m
than to 0.90 $\mu$m. Therefore, the observational evidence points
to a possible Cr-rich basaltic composition on the surfaces of
(7472) Kumakiri and (10537) 1991 RY16.

Concerning the dynamical behavior of these two asteroids, Table
\ref{proper} lists their proper elements and diameters, as well as
those of (1459) Magnya. The three asteroids are too small to be
differentiated bodies by themselves, they are quite spread in
proper elements space and do not belong to any of the asteroid
dynamical families identified in the outer belt. Therefore, they
are likely to be fragments from more than one differentiated
parent bodies. Nevertheless, at variance with (1459) Magnya,
(7472) Kumakiri and (10537) 1991 RY16 evolve very close to the non
linear secular resonance defined by the combination
$g_{0}+s_{0}-g_{5}-s_{7}\simeq 0$, where $g_i$ and $s_i$ represent
the frequencies of the perihelion $\varpi$ and node $\Omega$,
respectively ($i=0$ for asteroid, $i=5$ for Jupiter, $i=7$ for
Uranus; see Milani \& Kne\v{z}evi\'c, 1992). A 50 My simulation of
the orbits of these two asteroids, including gravitational
perturbations from the four major planets, indicate that they have
quite stable orbits showing a slow circulation of the angle
$\varpi_{0}+\Omega_{0}-\varpi_{5}-\Omega_{7}$. Although this may
be just a coincidence, a dynamical connection between (7472)
Kumakiri and (10537) 1991 RY16 cannot be ruled out and should be
addressed by more detailed studies.

\section{Conclusions}

We presented visible spectroscopic observations of two asteroids,
(7472) Kumakiri and (10537) 1991 RY16, located in the outer belt.
The main goal of our work was to show that these observations are
compatible with the V-type taxonomic class. Therefore, these
bodies would constitute the second and third basaltic asteroids
discovered up to now in that part of the Main Belt.

However, the presence of a shallow absorption band in the spectra
around 0.65 $\mu$m opens some questions about the actual
mineralogy of these two asteroids. This band is likely to be
related to the presence of Cr$^{3+}$ cations, and provides
evidence for a possible a Cr-rich basaltic surface.

The spectroscopic similarities among the two asteroids, together
with some shared dynamical properties, point to the idea of a
common origin from the breakup of a differentiated parent body in
the outer belt. Further studies, including near infrared (NIR)
spectroscopic observations, are mandatory to better address these
issues.

\begin{acknowledgements}
We thank Calar Alto Observatory for allocation of Director's
Discretionary Time to this programme. Fruitful discussions with D.
Nesvron\'{y} are also highly appreciated. Based on observations
collected at the Centro Astron\'omico Hispano Alem\'an (CAHA) at
Calar Alto, operated jointly by the Max- Planck Institut f\"{u}r
Astronomie and the Instituto de Astrof\'{\i}sica de Andaluc\'{\i}a
(CSIC). RD acknowledges financial support from the MEC (contract
Juan de la Cierva).
\end{acknowledgements}

%--------------------------------------------------
%

%
%--------------------------------------------------
%
\newpage
\begin{table*}
\caption{Observational circumstances for the targets: Universal
Time (UT), heliocentric distance ($r$), geocentric distance
($\Delta$), phase angle ($\phi$), visual magnitude ($V$), airmass
and total exposure time ($T_{\mathrm{exp}}$).} \label{observ}
\centering
\begin{tabular}{lccccccc}
\hline\hline \noalign{\smallskip}
 Asteroid & UT & $r$ [AU] & $\Delta$ [AU] & $\phi$ [deg] & $V$ [mag]& airmass & $T_{exp}$\\
\hline \noalign{\smallskip}
November 14 &  & & & & & & \\
(7472) Kumakiri & 03:25:00 & 2.920 & 2.123 & 13.5 & 16.6& 1.037 & 3400 sec\\
(10537) 1991 RY16 & 00:05:08 & 3.040 & 2.053 & 1.8 & 17.1 & 1.091 & 4000 sec\\
\noalign{\smallskip} \hline \noalign{\smallskip}
December 29 & & & & & & & \\
(7472) Kumakiri & 21:48:52 & 2.873 & 1.901 & 3.8 & 15.9 & 1.094 & 3600 sec\\
\noalign{\smallskip} \hline
\end{tabular}
\end{table*}
%
%--------------------------------------------------
\newpage
\begin{table*}
\caption{Proper elements and sizes of V-type asteroids in the
outer belt. For (1459) Magnya, the last column gives the diameter
from Delbo et al. (2006). For (7472) Kumakiri and (10537) 1991
RY16, the diameter was computed assuming an albedo of 0.40.}
\label{proper} \centering
\begin{tabular}{ccccc}
\hline\hline \noalign{\smallskip}
Asteroid & $a_{p}$ [AU]& $e_{p}$& $\sin I_{p}$ & $D$ [km] \\
\hline \noalign{\smallskip} \\
(1459) Magnya & 3.14986& 0.2183 & 0.2651& $17.0 \pm 1.0$\\
(7472) Kumakiri & 3.01033& 0.1372 & 0.1562 & 8.5\\
(10537) 1991 RY16 & 2.84958 & 0.1023& 0.1101 & 7.3\\
\hline
\end{tabular}
\end{table*}
%
%--------------------------------------------------
\newpage
%
%--------------------------------------------------
\newpage

%
%__________________________________________________________________
\newpage
%
%__________________________________________________________________
%

\end{document}